\documentclass[aip,amsmath,reprint,jcp]{revtex4-1}
\usepackage{docs}%
\usepackage{bm}%
\usepackage{graphicx}
\usepackage{dcolumn}% !!!package missed to have the right style for caption
\expandafter\ifx\csname package@font\endcsname\relax\else
 \expandafter\expandafter
 \expandafter\usepackage
 \expandafter\expandafter
 \expandafter{\csname package@font\endcsname}%
\fi
\begin{document}

\title{Metallic nanoparticles meet Metadynamics}%

\author{L. Pavan, K. Rossi and F. Baletto}%
\email{francesca.baletto@kcl.ac.uk}
\affiliation{Physics Department, King's College London, WC2R 2LS, UK}%

\date{July 2015}%
\revised{}%
\begin{abstract}
We show how standard Metadynamics coupled with classical Molecular Dynamics can be successfully applied to sample  the configurational and free energy space of metallic and bimetallic nanopclusters via the implementation of collective variables related to the pair distance distribution function of the nanoparticle itself. As paradigmatic examples we show an application of our methodology to Ag$_{147}$, Pt$_{147}$ and their alloy Ag$_{shell}$Pt$_{core}$ at 1:1 and 2:1 chemical compositions. The  proposed scheme is not only able to reproduce known structural transformation pathways, as the five and the six square-diamond mechanisms both in pure and core-shell nanoparticles but also to predict a new route connecting icosahedron to anti-cuboctahedron.
\end{abstract}
\maketitle

\section{Introduction}
Mono- and bi-metallic nanoparticles (mNPs) find a wide number of applications ranging from catalysis and biomedicine to optoelectronics and magnetic data storage due to their high surface to volume ratio, anisotropy and $d$-band shift. \cite{Ferrando2008,Baletto2005} Nanoclusters' chemophysical properties in fact strongly depend on the interplay between their size, morphology, and chemical composition. Understanding the relative stability of a configuration depends not only on weather it is a local minimum on the potential energy landscape but also on complex entropic contribution difficult to address experimentally \cite{Wales2014}. An estimate of the magnitude of free energy barrier and insights on the mechanisms involved in solid-solid structural rearrangements among different minima of the free energy landscape may shed light on how mNPs chemical features vary due to ageing and external factors such as temperature or pressure. 

A variety of theoretical attempts have been presented in the literature to extend classical thermodynamics concepts to fully address and include the contributions deriving by the large presence of surface atoms.\cite{Barnard2004,Berry2005,Li2014a} Two are the general approaches to obtain a quantitative sampling of the nanoclusters' free energy surface (FES) via computational techniques. Double ended searches are based upon the foreknowledge of the initial and final point of the transition and consist in analysing accurately the transition networks between the two target configurations, notable examples are the string method \cite{E2002}, the discrete path sampling \cite{Wales*2004} and the nudged elastic band \cite{Henkelman2000}. Temperature accelerated and biased sampling techniques represent instead an opposite approach: an initial configuration of the system is excited or perturbed and forced to visit new and initially unknown attraction basins. Adaptive biasing force \cite{Darve2008}, umbrella sampling \cite{Torrie1977} and parallel tempering \cite{Earl2005} are all renown techniques based upon this idea. Both methodologies have found a wide application in the analysis and study of structural transformations in nanoclusters. \cite{Adjanor2006,Neirotti2000,Calvo2011,Calvo2012,Calvo2014,Noya2006} Double ended methods weakness lies in the curse of dimensionality which afflicts the potential energy surface (PES) minima basins network such that these algorithms need a larger and larger amount of computational expenses with increasing mNP size. Instead perturbation methods have been commonly used to detect order-disorder transition while rarely solid-solid ones have been characterized or simulated by these.

In this paper we will show for the first time how Metadynamics can be successfully employed to sample the configurational space of relatively large metallic and bimetallic nanoparticles at finite temperatures and how the free energy barrier for various minimum-to-minimum can be estimated. 
Metadynamics (MetaD) algorithm combines the ionic dynamics with a history dependent potential exploited to accelerating rare events and to reconstruct the free energy surface projections into an order parameters, named collective variables (CVs), space. \cite{Laio2002} First introduced in the biophysics community to elucidate ligand binding and protein folding phenomena \cite{Laio2008}; it has been demonstrated in the literature that this method can be also applied for finite inorganic nanosystems such as semiconductor/quantum dots nanoparticles \cite{Bealing2009,Bealing2010}, alkali halides nanostructures \cite{Bochicchio,Liu2011} Lennard-Jones \cite{sprint,Tribello2010} or metallic nanoparticles \cite{Santarossa2010,Pavan2013} of a very small size. However, it has been never applied extensively and systematically for metallic nanoclusters larger than few tens of atoms nor to bimetallic cases. We device molecular dynamics (MD), based on the second-moment approximation of the tight-binding potential (TBSMA), coupled to MetaD as a cost effective and accurate technique. Our MetaD scheme, with CVs based on the pair distribution function, has shown to be able to reproduced the well known five and six Diamond-Square-Diamond (DSD) \cite{Lipscomb1966a} mechanism and for the first time the interconversion of an icosahedron into a anti-cuboctahedron is reproduced validating Mackay's theoretical speculation. \cite{Mackay1962}

%More in depth, this enhanced sampling algorithm is based upon the coarse graining of the multidimensional energy landscape of the system into a set of sensibly chosen collective variables (CV) along which a repulsive history-dependent biasing potential is employed to overcame energy barriers and discourage the permanence on already visited conformational minima. 

\section{Models and Method}
\begin{figure}[h!]
\begin{center}
\includegraphics[width=7cm]{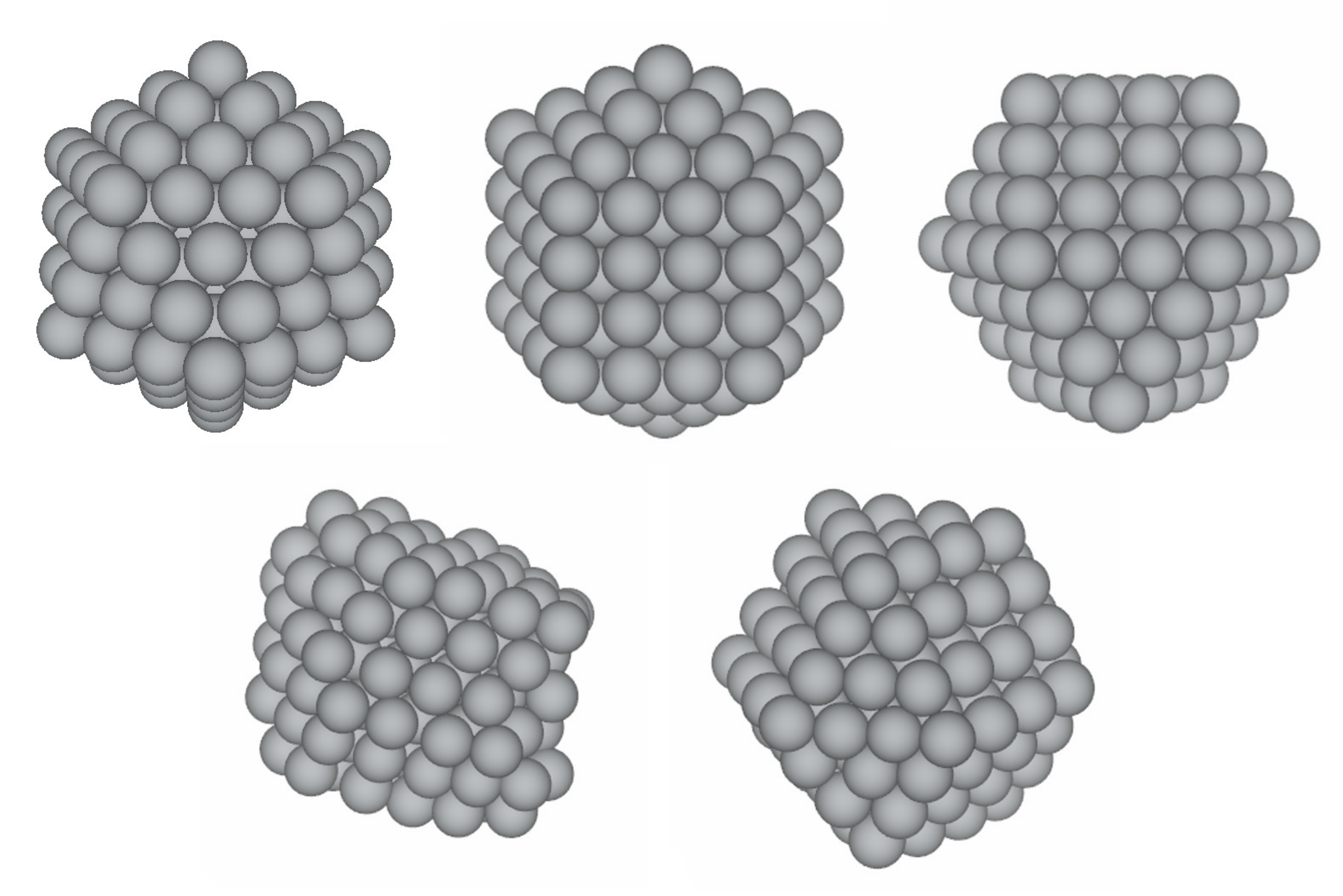}
\caption{Closed shell polyhedra at 147 atoms. Top row, from left to right: icosahedron, decahedron, cuboctahedron. Bottom row: hexagonal closed packed and anticuboctahedron.}
\label{Fig:collective}
\end{center}
\end{figure}

We consider monometallic (Ag,Pt) and bimetallic (AgPt) nanoparticles at 1:1 and 2:1 chemical compositions at a selected size of 147 atoms. At this size, noble and quasi-noble metals adopt geometrical closed shell polyhedra such as icosahedron (Ih), decahedron (Dh), truncated octahedron or cubo-octahedron (Co) and rarely hexagonal close-packed geometries (hcp), as reported in Figure \ref{Fig:collective}. As shown in details by Paz-Borbon et al. \cite{Paz-Borbon2008a}, AgPt has a small size mismatch, with a tendency to segregation and a core-shell chemical ordering with silver in the uppermost layer is usually preferred. 

We perform classical molecular dynamics simulations, coupled with MetaD, where a velocity-Verlet algorithm is used for solving the Newton's equations with a time step of 5 fs. An Anderesen thermostat is applied. Atomic interactions are modelled within the second moment approximation of the tight binding theory \cite{Rosato89} (TBSMA), where the potential $V_{TBSMA}(x)$ is the sum of atomic contribution, $E_{TBSMA}^i$, 
\begin{eqnarray}
E_{TBSMA}^i = && \sum_{j\ne i}^{n_v} A_{ab}e^{-p_{ab}\left(\frac{r_{ij}}{r^0_{ab}} - 1 \right)} \\ \nonumber
&& -\sqrt {\sum_{j\ne i}^{n_v} \xi^2_{ab}e^{-2q_{ab}\left(\frac{r_{ij}}{r^0_{ab}} - 1\right)} }  \mbox{~~~~~~.}
\end{eqnarray}

Here the sum is extended up to the number of atoms $n_v$ within an appropriate cut-off distance; $a$ and $b$ refer to the chemical species of the two atoms, $r_{ab}^0$ is the bulk nearest neighbour distance, $A_{ab}$ and $\xi_{ab}$ are related to the cohesive energy, while $p_{ab}$ and $q_{ab}$ and their product determine the range of the repulsive and of the attractive part of the potential. All the four parameters have been fitted using bulk properties. Parametrization of Ag and Pt interaction follows what suggested in Ref. \cite{Cleri1993} while for AgPt parameters are fitted according to Ref.\cite{Paz-Borbon2008a}. 

MetaD enhanced sampling algorithm is based upon the construction of a history-developing biasing potential, $\Delta V$, evolving as the sum of gaussians of height $\omega$ and width ${\sigma}$ added every ${t_{G}}$ time interval,
\begin{eqnarray}
 V(x) &=& V_{TBSMA}(x) + \Delta V\left(S(x),t\right) \\ \nonumber
 \Delta V\left(S(x),t \right) &=& \sum_{t' = t_{G}, 2t_{G}..} \omega e^{-\frac{ \left[S(x)-s(t') \right]^{2}}{2\sigma^{2}} }  \mbox{~~~.}
 \label{eq:met}
\end{eqnarray}
$S(x)$ represents a set of collective variables and defines the order parameter space on which the added gaussians lie; $s(t')$ is the instanteneous value of the collective variable. When the collective variables have sampled all the conformation's space they become fully diffusive and free energy along this order parameters can be reconstructed as the negative of the meta-potential $\Delta V(S(x),t)$. The efficiency and physical faithfulness of MetaD is strongly related to the choice of a sensible set of collective variables. The chosen metric should be able to distinguish between the various configurations the system visited during a run and the CVs should also be representative of all slow degree of motions involved in a structural transformation. Furthermore their number should be limited in such a way that we have a low dimensionality space to explore avoiding undesired computational expenses. If one of the above conditions is not respected the system can be forced to visit high energy regions and/or attraction basin of interest may be hidden. \cite{Laio2008}

\begin{figure}[h!]
\begin{center}
\includegraphics[width=8.2cm]{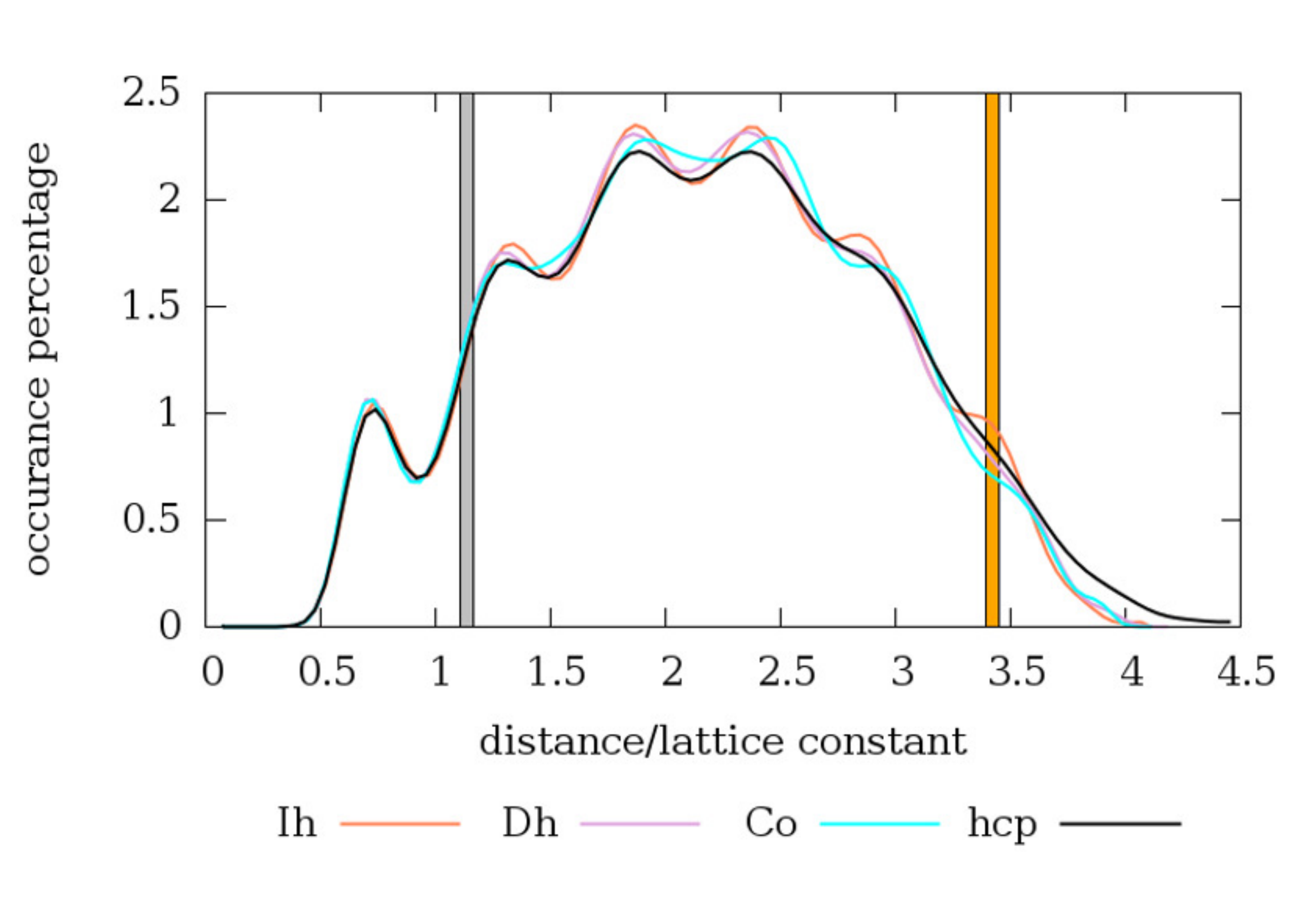}
\caption{Pair distance distribution function for icosahedral (green), decahedral (blue), cuboctahedral (red), and hcp (black) motifs of a 147 atoms cluster. The selected windows around the stacking fault number (SFN), and the maximum pair distribution difference (MPDD), are highlighted  in grey and orange shadowing, respectively.}
\label{Fig:PDF}
\end{center}
\end{figure}

The research of a good order parameter able to differentiate among the complex configuration space of a cluster is, however, not trivial \cite{Wales2015,Pietrucci2015}. On the other hand, an almost complete information on NP morphology is encoded in its pair distance distribution function (PDDF), depicted in Figure \ref{Fig:PDF}, where the four isomers shown in Figure \ref{Fig:collective} are considered.

Hence we broaden the restriction on our CV set according to the following criterion: it must able to clearly distinguish out the above mentioned geometries as these are the main structures of interest in our research. Thus, with a low computational cost and only a small loss in accuracy due to a partial correlation of the two CVs, we introduce specifically tailored order parameters corresponding to window function, $WF$, on the PDDF constructed via a sigmoid function:

\begin{eqnarray}
WF(x) = \sum_{i,j; i \neq j} \frac{1 - \left(\frac{r_{ij} - d_{0}}{r_{0}} \right)^{6}}{1 - \left(\frac{r_{ij} - d_{0}}{r_{0}} \right)^{12}} \mbox{,}
\label{eq:cv}
\end{eqnarray}
where $r_{ij}, r_{0}, d_{0} $ are respectively the distance between atom $i$ and atom $j$, $r_{0}$ the window width, always set to 0.05 of the bulk lattice constant, and $d_{0}$ is the characteristic distance. $d_0$ is set to be 1.354 and 3.4 of the bulk FCC lattice parameter respectively for the stacking fault number (SFN) and the maximum pair distance difference (MPDD), as shown in Figure \ref{Fig:PDF}.

A qualitative argument can be provided to support our collective variable choice. The MPDD is chosen to be set where the difference between configurations in the pair distribution function has a maximum. In such a way we are confident of being able to discriminate at least the main structural topologies. The stacking fault number instead is related to a characteristic hcp peak in the radial distribution function. Physically it is a topological defect obtained due to the intersection of two planes with different symmetry orientation. At the nanoscale, as in the bulk, phase transformations happen via stacking faults, as demonstared for the diamond-square-diamond mechanisms. \cite{Uppenbrink1991}
Consequently a bias on this CV should not constrain the free energy surface exploration to unphysical structural transition or highly energetic states.  

We resort to the use of these two window functions on specific lattice distances because they adopt different values for each of the geometries of interest in our research and are computationally very cheap, being two body terms their calculations scales as $N^{2}$. This will enable us to easily extend our methodology to larger systems. On the other hand, this set of CVs can be applied with a partial confidence to bimetallic nanoalloys. The main problem lies in the fact that the presence of different chemical species is not treated explicitly by the earlier on presented window functions, hence structural transformations involving chemical reordering are rarely reproduced biasing on those coordinates. However, due to the fact that Ag and Pt have a small size mismatch, and because we focus our attention on the transformation of a core-shell ordering, SFN and MPDD still capture the geometrical structural transition pathways of interest.

\subsection{Sampling and reconstructing the FES of a metallic nanoparticle}

\begin{figure}[!h]
\begin{center}
\includegraphics[height =2.8cm]{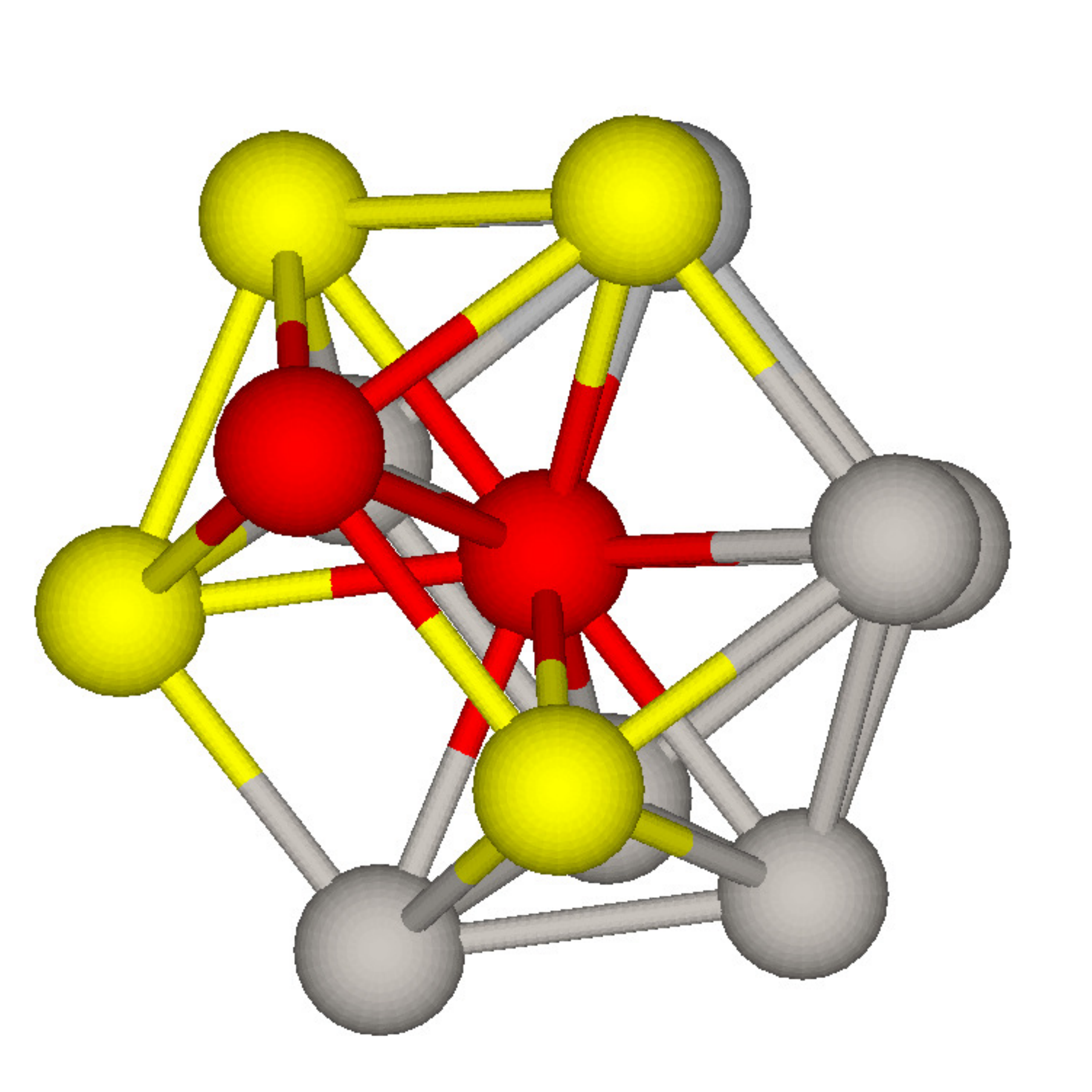}
\includegraphics[height =2.8cm]{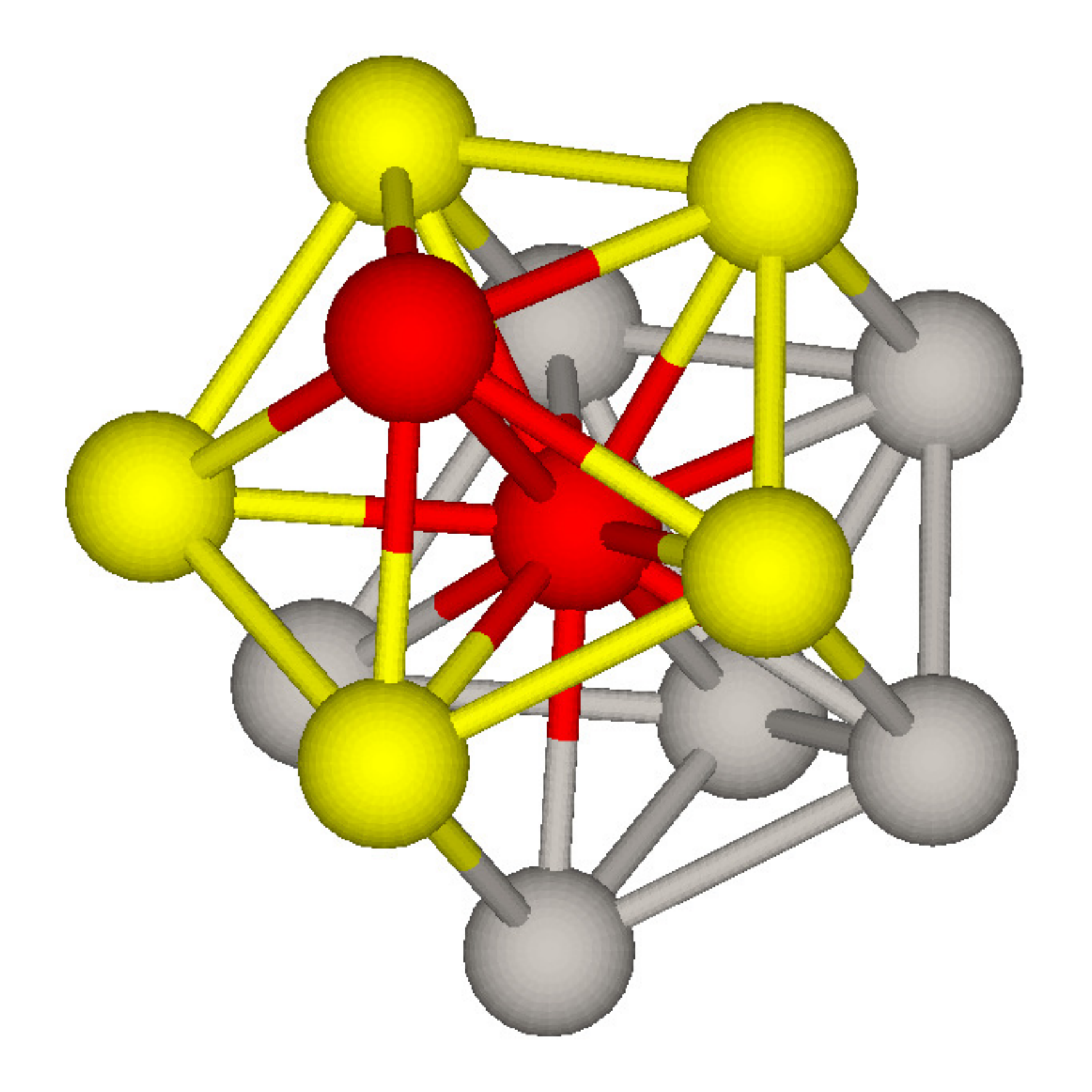}
\caption{Common neighbourhood, in yellow, for the red-coloured atomic pair. Left panel shows a (422) signature, associated with twin boundaries, while the right panel depicts a (555) signature, which is related to pairs lying of a 5-fold symmetry axis.}
\label{Fig:CNA}
\end{center}
\end{figure}

During the course of the simulations the cluster topology is monitored on-the-fly during a MetaD-MD simulation via common neighbour analysis (CNA)\cite{Honeycutt1987}. The CNA determines the local environment of all the nearest neighbours atom pair and produces a number of signatures percentages that characterises the whole cluster. A CNA signature is made by three integers $(r,s,t)$: $r$ is the number of common nearest neighbour, $s$ the number of bonds between the $r$-common nearest neighbours and $t$ is the longest chain among the $s$-bonds. The percentage of each signature corresponds to features of the considered NP. 
Different CNA signatures are able to distinguish whether a pair of atoms is in a bulk environment or on the surface, whether the bulk is crystallographic or not, and if the pair belongs to a symmetry axis. A fast NP geometry taxonomy may follow looking at the (4,2,2) and the (5,5,5) signatures. Their local neighbourhood is depicted in Figure \ref{Fig:CNA}. The (4,2,2) varies between 39\% (Ih) to 25\% (Dh and hcp) and zero (Co); while the (5,5,5) ranges between 5.2 \% (Ih), 0.9 \% (Dh) and zero (Co and hcp).  

\begin{figure}[h!]
\includegraphics[width=8.5cm,height=14cm]{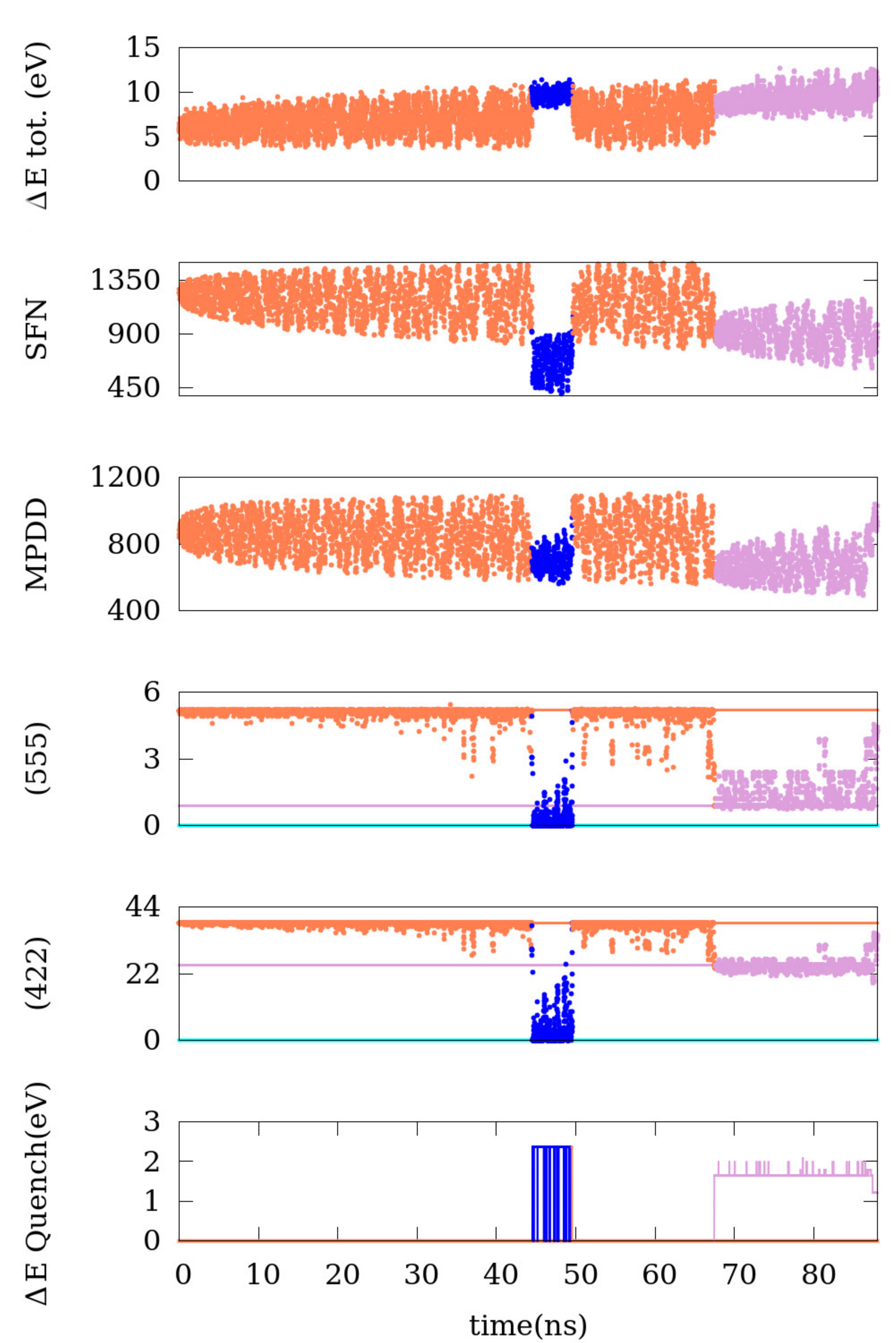}
\caption{$Ag_{shell}Pt_{core}$ MetaD-MD simulation outcome and analysis. From top to bottom: Total energy difference ($\Delta E_{tot} = E_{tot} - E_{Ih}^{quench}$ in eV) and CV values are shown in the first three panels. In the lowest three panels, on-the-fly CNA analysis combined with a fast quenching for reproducing the map of all the basin encountered ($\Delta E_{quench} = E_{quench} - E_{Ih}^{quench}$.  Reference energy corresponds to the one of a quenched Ih $E_{Ih}^{quench}$. A colour scheme is applied for remarking all the independent basin/shape assumed: Co-basin in blue, Ih-basin in orange, Dh-basin in pink.}
\label{Fig:analysis}
\end{figure}

A significant change in the collective variables and in the CNA signatures is therefore distinctive of structural transition, as in the paradigmatic example of an Ag$_{92}$@Pt$_{55}$ run sketched in Figure \ref{Fig:analysis}, where three different basins, Ih (orange), Co (blue) and Dh (pink), are explored. Furthermore structures close to any presumed phase change are quenched in order to identify precisely the initial and final shape of the transition on the potential energy surface. The PES profile obtained after that fast quenching is reported in the lower panel of Figure \ref{Fig:analysis}. 

As it may happen that the diffusive regime is not reached, the free energy reconstruction is not merely the inverse of the added MetaD potential, $\Delta V$ in Eq. \ref{eq:met}. To obtain an overestimate of the free energy barrier separating two minima, we use the above on-the-fly analysis in order to monitor and collect after how many Gaussian depositions a transition takes place. Thus, an upper bound estimate of the free energy profile separating the two minima can be calculated as the negative of the repulsive MetaD potential deposited on the initial basin and needed to visit the following conformational minimum. 

In order to validate our approach and to compare our sampled free energy landscape with the corresponding potential energy one, the activation energy on the PES is calculated by means of a DETPS \cite{Wales*2004} where the relaxed structures obtained from the fast quenching of the configurations visited during a MetaD run serve as the initial and endpoint of the transformation pathway, respectively.

\section{Results}

Gaussians 0.25 eV high are deposited with a period of 20ps. The Gaussian width ${\sigma}$ is 15 along the SFN dimension and 10 for the MPDD. Such values are based on the standard deviations of the CVs during unbiased molecular dynamics simulations at 300 K and are chosen to assure the nearly spherical shape of the energy wells and thus an optimal conformational flooding. All the following results are obtained at room temperature of 300 K.

\begin{figure}[!ht]
\includegraphics[width=8.5cm, height=3cm]{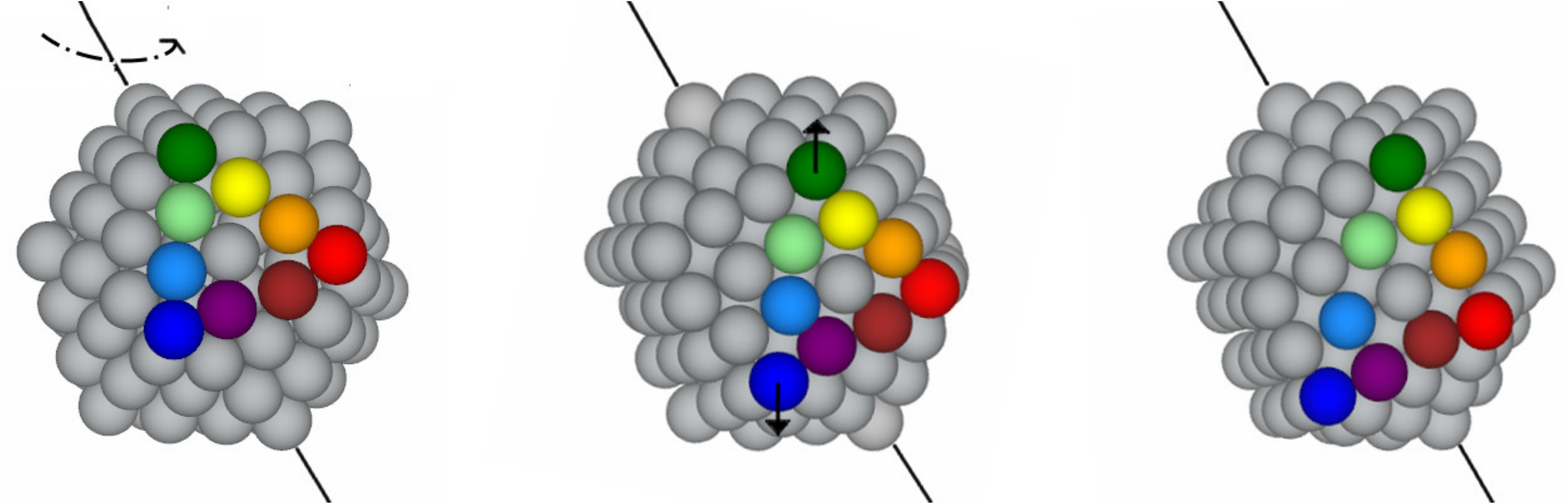}
\includegraphics[width=8.5cm, height=3cm]{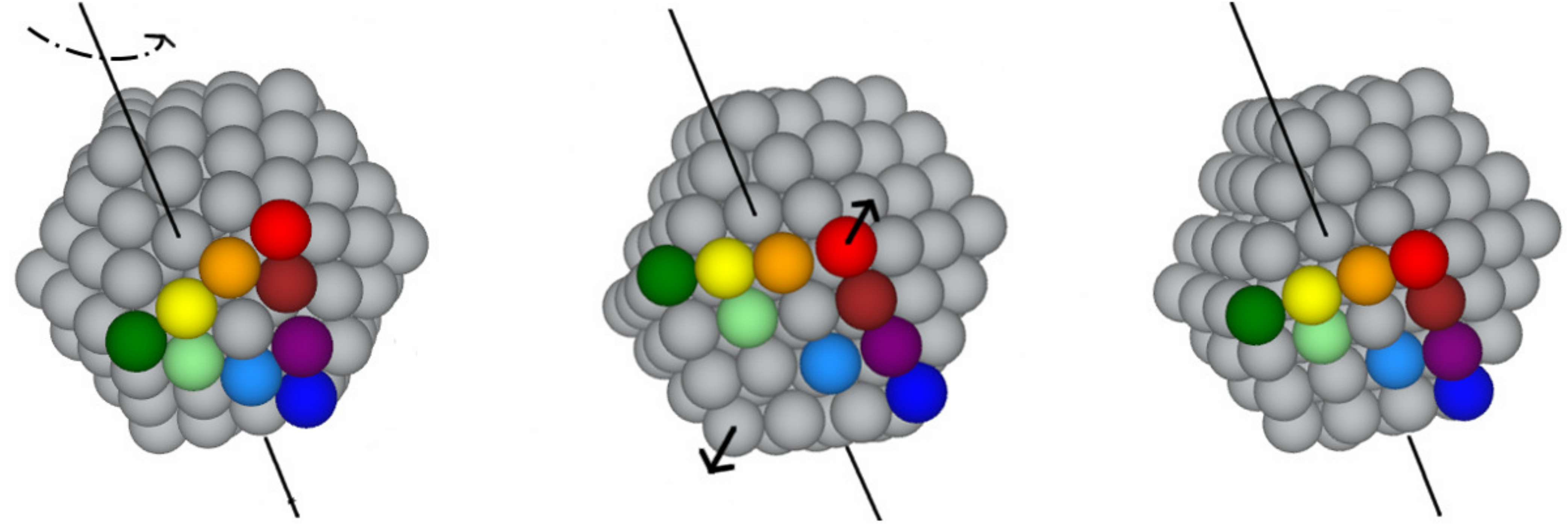}
\caption{Initial, saddle and final configurations of DSD mechanism for Ih into Co (top row) and Ih into Dh (bottom row) transformation. Multicolored atoms delimit a facet in the original Ih. Rotation axis is shown, }
\label{Fig:path}
\end{figure}
\begin{figure}[!hb]
\includegraphics[width=8.5cm, height=3cm]{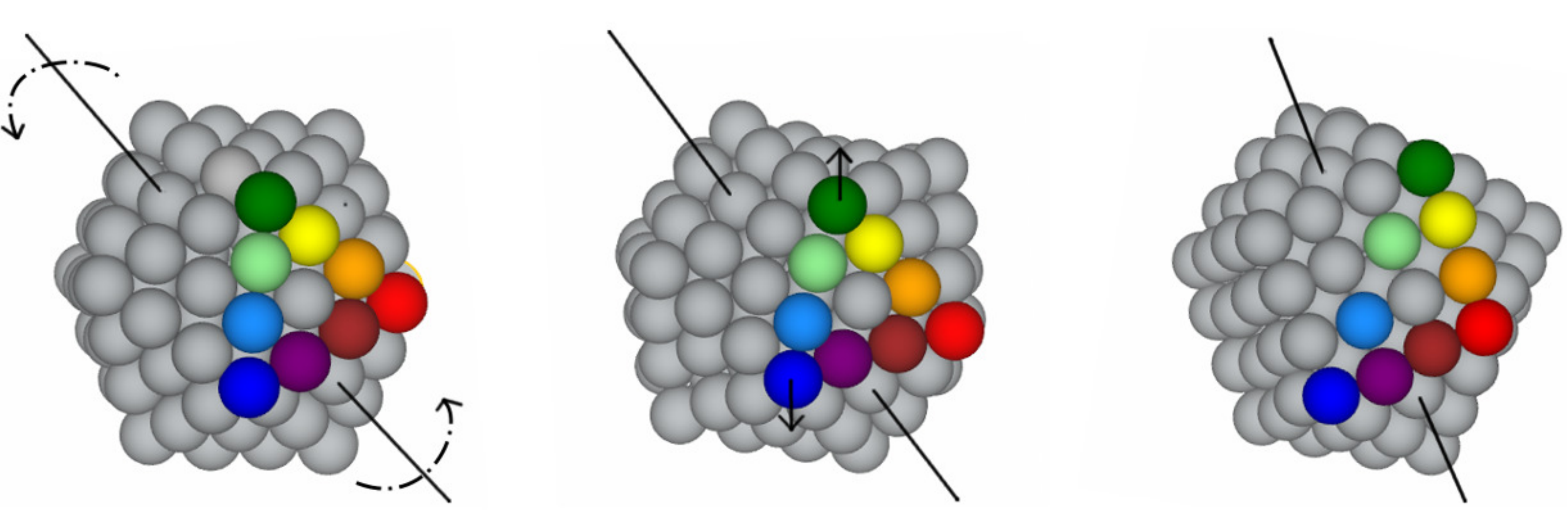}
\caption{Ih transforming into aCo via DSD mechanism. Saddle point is pictured in between the two. Multicolored atoms delimit a facet in the original icosahedron which first rotate by 60 in opposite direction with respect to the parallel triangular facets and becomes part of a diamond facet which then transforms into a square.}
\label{Fig:acoico}
\end{figure}

The detected structural transformations in the Ag clusters are Dh into Ih, Co into Ih and backwards, Ih into aCo and backwards. In the case of the Pt cluster only transformations into Ih geometry from Dh and Co have been observed before the exploration of highly defected structures.
The general solid-solid structural transformation mechanism between the above mentioned geometries is the so-called diamond - square - diamond (DSD) process as shown in Figures \ref{Fig:path} and \ref{Fig:acoico}, where the motion of each atom is clearly identified via the different coloring. It consists of a stretching and rotation of triangular facets into a square by a collective dislocation of atoms along the surfaces involved. This dislocation corresponds to a rotation of different angles according to the initial and final configuration. Conversely a squared facet can transform into two triangular facets by the opposite movement. Five parallelograms are involved in this collective rearrangement in the case of the Dh-Ih transformation, and six in the case of the Co-Ih transformation, Figure \ref{Fig:path}.

As first noticed by Mackay \cite{Mackay1962} an anticuboctahedron could be transformed into an Ih throughout a rotation in opposite ways of the two triangular facets perpendicular to the aCo three-fold rotation axis, by 60 degrees with reference to each other, along the same axis. The transformation is similar with respect to the one of the Co, involving six parallelograms, however in this case two opposite parallel triangular facets rotate about their normal and three pairs of abutting triangular facets remain unchanged and rotate about the axis that is perpendicular to their common edge and belongs to the twin plane, see Figure \ref{Fig:acoico}.

\begin{figure}[!ht]
\begin{center}
\includegraphics[width=8.5cm, height=3cm]{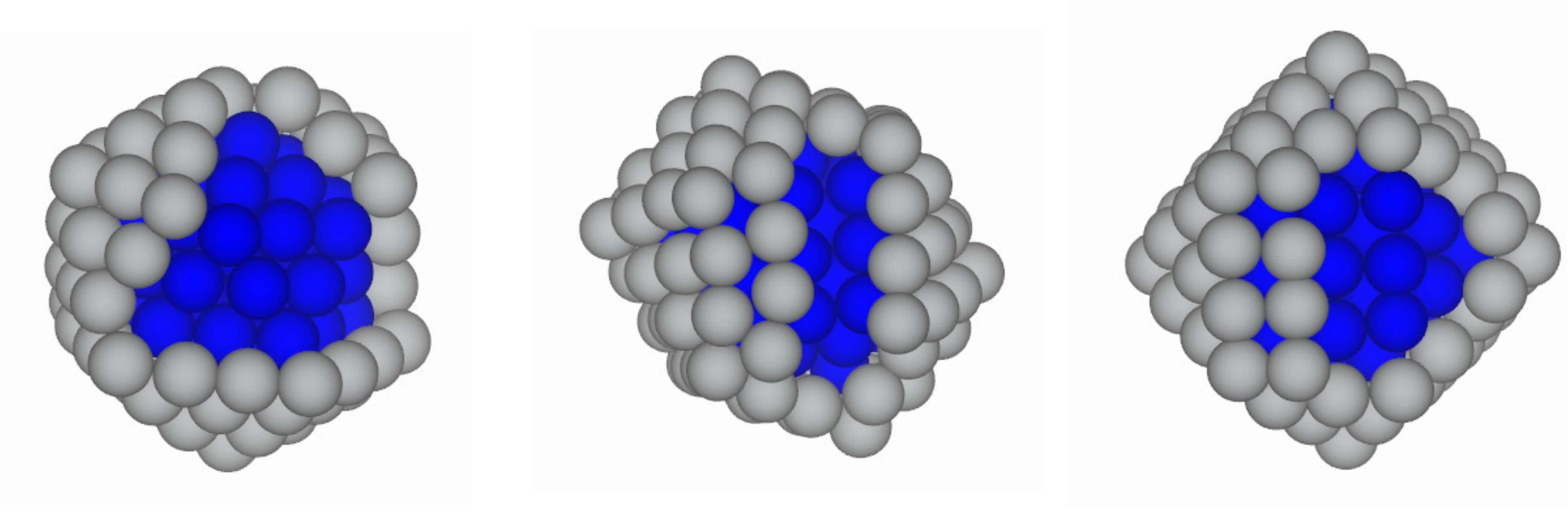}
\includegraphics[width=8.5cm, height=3cm]{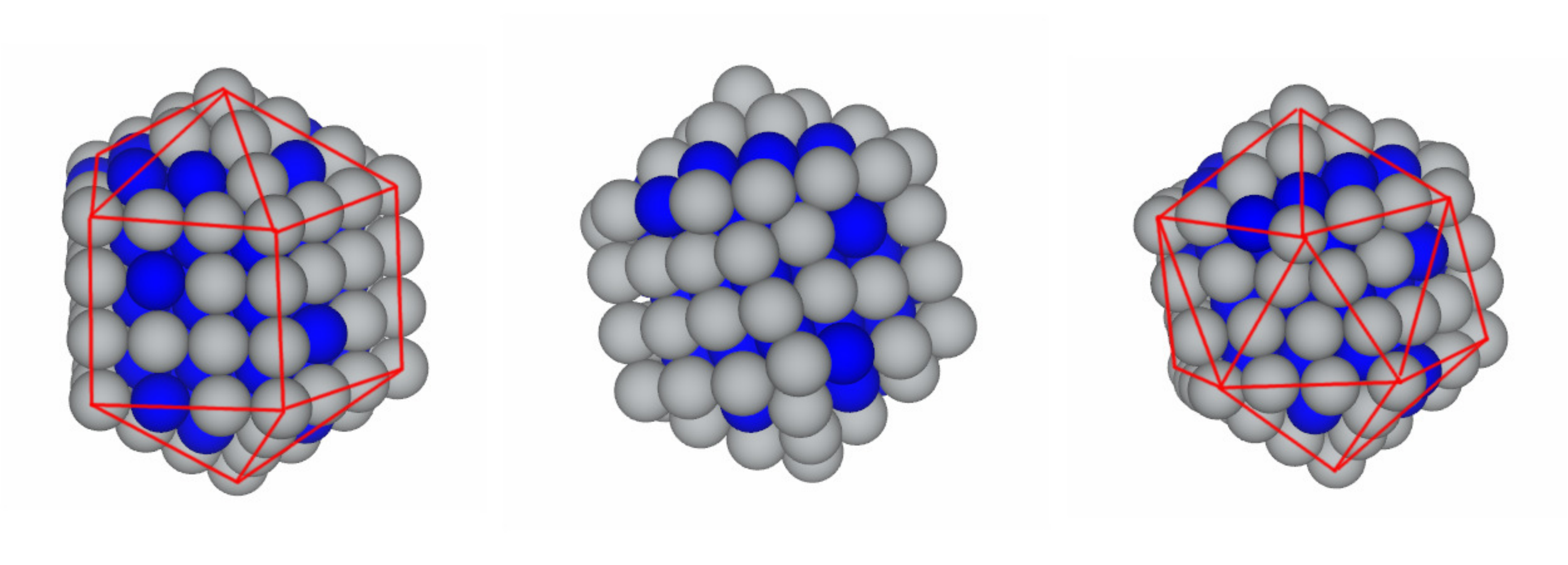}
\caption{Initial, saddle and final configuration in the case of AgPt systems. Top row: Ih into Co transformation in Ag$_{92}$@Pt$_{55}$. Some surface atoms are removed from the picture in order to show that the transformation is simultaneous in the two chemical species. Bottom row: Dh into Ih transformation in Ag$_{74}$Pt$_{73}$. Ag atoms are in grey and Pt in blue.  Red lines highlight facets involved in the DSD mechanism.}
\label{Fig:bimdsd}
\end{center}
\end{figure}
\begin{figure}[!h]
\begin{center}
\includegraphics[width=8.5cm]{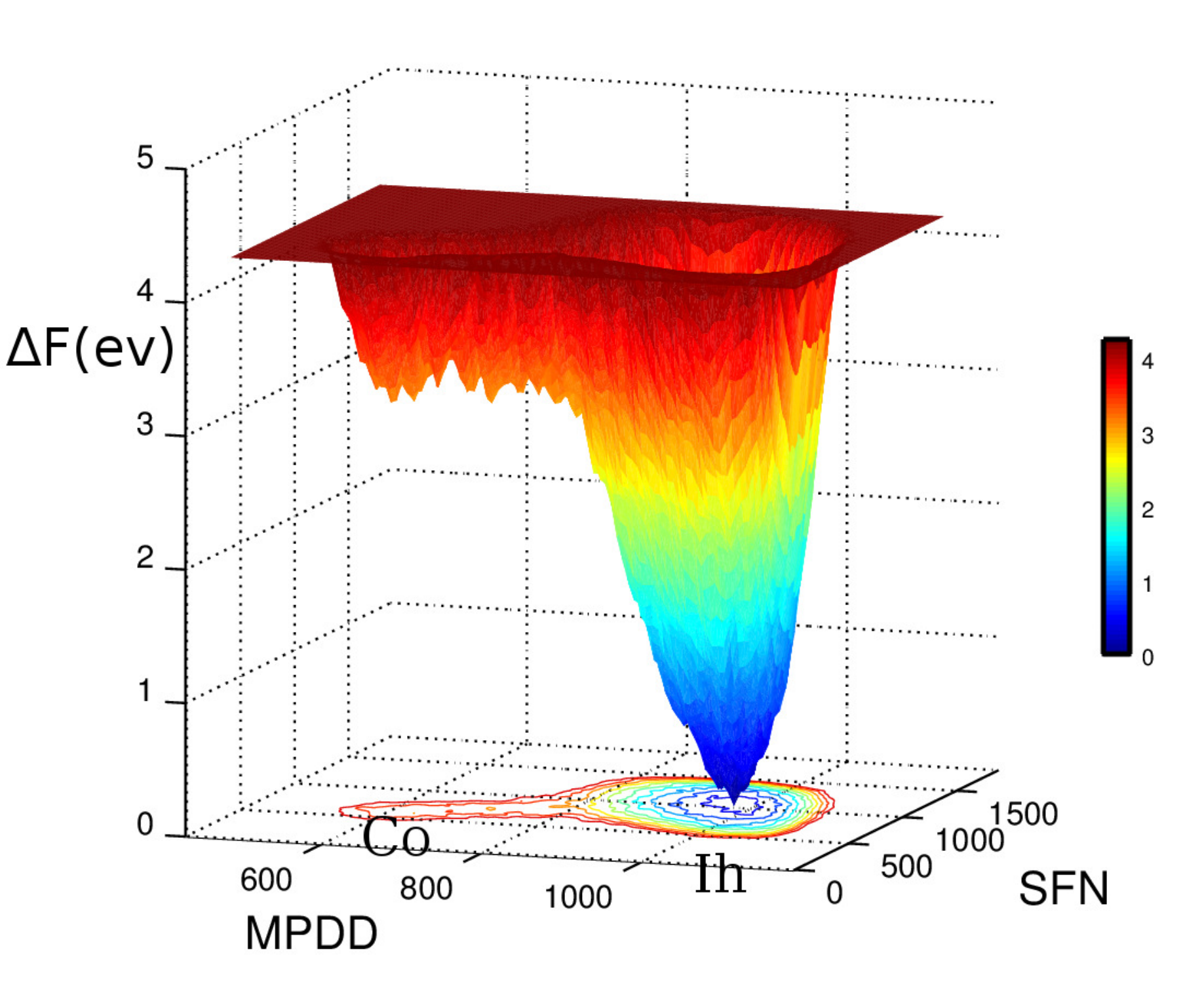}
\caption{Free energy surface reconstruction for an Ag$_{92}$@Pt$_{55}$ run where a diffusive regime in the exploration of Co and Ih basins is found. Color key helps to visualize free energy differences, $\Delta F$, with respect to $E_{Ih}^{quench}$ at a given point in the CV space.}
\label{Fig:FES}
\end{center}
\end{figure}

The simulation parameters used for the monometallic systems are also fruitfully adopted to investigate structural transformations in AgPt NPs. In the case of a Ag$_{92}$@Pt$_{55}$ -where "@" stands for a shell/core ordering. We would like to note that the results showed in Figure \ref{Fig:analysis} refers to the same system. The pathways followed during the structural transitions are the same described early. The transition happens simultaneously in the two chemical species as shown in the top row of Figure \ref{Fig:bimdsd}. Ih configuration is clearly more energetically favourable with respect to the Co one. The collective twisting of two triangular facets is energetically expensive due to the strain increase while the almost barrierless transformation of Co into Ih geometry is corroborated by Figure \ref{Fig:analysis} bottom panel. Roughness of the landscape can be bestowed to the sampling of intermediate defected structures. Ag$_{92}$@Pt$_{55}$ reaches a diffusive regime between Ih, Dh and Co configurations before visiting higher energy conformations, allowing a FES reconstruction accordingly to Ref.\cite{Laio2002,Laio2008}. The Ih to Co FES reconstruction is shown in Figure \ref{Fig:FES}.

A quantitative analysis of the energy barriers defining the detected processes was done exploiting our on-the-fly analysis tool also when no diffusive regime in the conformation's space exploration was found. In that scenario, the free energy barrier is calculated via the amount of gaussians deposited in a simulation starting from specific initial structures according to the initial point of the transition of interest. Results for the monometallic and bimetallic transformations are resumed in Table \ref{table:tab} together with the DETPS results in parenthesis. We would like to note that the geometric structural transition pathways, including the saddle configuration, on the PES and on the FES are almost identical. 
\begin{table}[!h]
\caption{Based on a TBSMA molecular dynamics, free energy barrier by MetaD at 300 K, and the activation energy calculated by DEPTS, in brackets, for Ag$_{147}$, Pt$_{147}$ and Ag$_{92}$@Pt$_{55}$, respectively. "NA" means that that pathway has not been observed in MetaD simulations. All the values are in eV.}
\begin{large}
\begin{tabular}{l|c|c|c}
\hline 
Mechanism & Ag & Pt & Ag$_{92}$@Pt$_{55}$ \\ 
\hline
\hline
Dh $\rightarrow$ Ih & 0.6 (0.45) & 1.9 (2.11) & 1.5 \\

Ih $\rightarrow$ Dh & NA   (2.71) &  NA   (4.55) & 4.3\\
\hline
Co $\rightarrow$ Ih & 0.4 (0.5) & 1.7 (1.9) & 0.25\\

Ih $\rightarrow$ Co & 3.2 (3.2) &  NA (5.08) & 3.6\\
\hline
aCo $\rightarrow$Ih & 0.5 (0.64) & NA    (1.87) & NA \\

Ih $\rightarrow$ aCo & 3.0 (3.49) &  NA  (5.13) & NA\\
\hline
\hline
\end{tabular}
\end{large}
\label{table:tab}
\end{table}

Free energy barriers heights can be related to the activation barrier for the diffusion of a Pt and Ag atom over Pt and Ag (111) and (100) surfaces. We observe that the energy barrier for processes identifies for the alloyed core-shell configurations is in between the one found for pure Ag and pure Pt clusters while Pt generally shows the highest free energy barriers and Ag the lowest. 

Finally, a 1:1 chemical composition, namely Ag$_{74}$Pt$_{73}$, with a core of Pt and a shell Ag-rich, has been considered. In this case, a Dh transformation into Ih and Ih into defected Co for is detected involving both a diamond-square-diamond and surface reconstruction mechanisms, as depicted in the bottom row of Figure \ref{Fig:bimdsd}. We would like to mention that in the case of mixed ordering, further studies are needed in order to understand whether the combination of these two structural transformation mechanisms represents the minimum energy pathway for the geometrical phase change of the examined structure.

\section{Conclusion}
We have shown that Metadynamics can be successfully adapted to the study of morphological transitions in metallic and bimetallic NPs by introducing specific collective variables. These are window function set onto the pair distance distribution function, and they refer to the stacking fault number and the maximum pair distribution distance. This set of collective variables has a wide range of applicability, from monometallic (e.g. Ag, Pt) to bimetallic systems with a small mismatch (e.g. AgPt). For nanoalloys with a large mismatch a limitation occurs. A broadening of the peaks in the PDDF may result in the impossibility of uniquely defining the characteristic distances, $d_0$ in Eq. \ref{eq:cv}. To recognising transitions occurring between two basins and to estimating the associated free energy barrier, even if no systematics diffusive regime may be obtained, we have introduced an on-the-fly analysis based on complementary geometrical and energetic order parameters, such as CNA and $\Delta E_{quench}$.
 
We have demonstrated that the solid-solid structural transitions among the most common closed shell polyhedra happen following the Lipscomb diamond-square-diamond mechanism, which is a collective screw dislocation motion. For the first time, we have shown that the Mackay's description of the interconversion of an Ih into an aCo throughout a DSD mechanism is reproducible. We remark that the DSD mechanism appears to be a universal pathway as it takes place in monometallic as well as in nanoalloys with a small either non negligible mismatch, such as in AgAu\cite{Chen2008} and AgPt.

Finally, we have demonstrated that the proposed MetaD scheme is able to predict free energy barriers in a very good agreement with DEPTS activation energy barriers. We have found that in bimetallic systems the free energy barriers lie roughly in between the values of the monometallic cases, although their numerical values are not simply rescaled.

\section*{ACKNOWLEDGMENTS}

LP and FB thank the financial support by U.K. research council EPSRC, under Grants No. EP/GO03146/1 and EP/J010812/1. KR acknowledges financial support by U.K. research council EPSRC, Grants No. ER/M506357/1. The simulations were carried out using the Faculty and Departmental computational facilities at King’s College London.

\bibliographystyle{jcp}
%\bibliography{library}

\begin{thebibliography}{10}

\bibitem{Ferrando2008}
{\sc R.~Ferrando}, {\sc J.~Jellinek}, and {\sc R.~L. Johnston},
\newblock {\em Chemical reviews} {\bf 108}, 845 (2008).

\bibitem{Baletto2005}
{\sc F.~Baletto} and {\sc R.~Ferrando},
\newblock {\em Reviews of Modern Physics} {\bf 77}, 371 (2005).

\bibitem{Wales2014}
{\sc D.~J. Wales} and {\sc P.~Salamon},
\newblock {\em Proc. Natl. Acad. Sci. U. S. A.} {\bf 111}, 617 (2014).

\bibitem{Barnard2004}
{\sc a.~S. Barnard} and {\sc P.~Zapol},
\newblock {\em J. Chem. Phys.} {\bf 121}, 4276 (2004).

\bibitem{Berry2005}
{\sc R.~S. Berry} and {\sc B.~M. Smirnov},
\newblock {\em Physics-Uspekhi} {\bf 48}, 345 (2005).

\bibitem{Li2014a}
{\sc Z.~H. Li} and {\sc D.~G. Truhlar},
\newblock {\em Chem. Sci.} {\bf 5}, 2605 (2014).

\bibitem{E2002}
{\sc W.~E}, {\sc W.~Ren}, and {\sc E.~Vanden-Eijnden},
\newblock {\em Physical Review B} {\bf 66}, 052301 (2002).

\bibitem{Wales*2004}
{\sc D.~J. {Wales *}},
\newblock {\em Molecular Physics} {\bf 102}, 891 (2004).

\bibitem{Henkelman2000}
{\sc G.~Henkelman} and {\sc H.~Jo},
\newblock {\em J. Chem. Phys.} {\bf 113}, 9901 (2000).

\bibitem{Darve2008}
{\sc E.~Darve}, {\sc D.~Rodr\'{\i}guez-G\'{o}mez}, and {\sc A.~Pohorille},
\newblock {\em The Journal of chemical physics} {\bf 128}, 144120 (2008).

\bibitem{Torrie1977}
{\sc G.~Torrie} and {\sc J.~Valleau},
\newblock {\em Journal of Computational Physics} {\bf 23}, 187 (1977).

\bibitem{Earl2005}
{\sc D.~J. Earl} and {\sc M.~W. Deem},
\newblock {\em Physical Chemistry Chemical Physics} {\bf 7}, 3910 (2005).

\bibitem{Adjanor2006}
{\sc G.~Adjanor}, {\sc M.~Ath\`{e}nes}, and {\sc F.~Calvo},
\newblock {\em The European Physical Journal B} {\bf 53}, 47 (2006).

\bibitem{Neirotti2000}
{\sc J.~P. Neirotti}, {\sc F.~Calvo}, {\sc D.~L. Freeman}, and {\sc J.~D.
  Doll},
\newblock {\em The Journal of Chemical Physics} {\bf 112}, 10340 (2000).

\bibitem{Calvo2011}
{\sc F.~Calvo} and {\sc C.~Mottet},
\newblock {\em Physical Review B} {\bf 84}, 035409 (2011).

\bibitem{Calvo2012}
{\sc F.~Calvo}, {\sc J.~P.~K. Doye}, and {\sc D.~J. Wales},
\newblock {\em Nanoscale} {\bf 4}, 1085 (2012).

\bibitem{Calvo2014}
{\sc F.~Calvo} and {\sc E.~Yurtsever},
\newblock {\em The Journal of chemical physics} {\bf 140}, 214301 (2014).

\bibitem{Noya2006}
{\sc E.~G. Noya} and {\sc J.~P.~K. Doye},
\newblock {\em The Journal of chemical physics} {\bf 124}, 104503 (2006).

\bibitem{Laio2002}
{\sc A.~Laio} and {\sc M.~Parrinello},
\newblock {\em Proceedings of the National Academy of Sciences of the United
  States of America} {\bf 99}, 12562 (2002).

\bibitem{Laio2008}
{\sc A.~Laio} and {\sc F.~L. Gervasio},
\newblock {\em Reports on Progress in Physics} {\bf 71}, 126601 (2008).

\bibitem{Bealing2009}
{\sc C.~Bealing}, {\sc R.~Marton\'{a}k}, and {\sc C.~Molteni},
\newblock {\em The Journal of chemical physics} {\bf 130}, 124712 (2009).

\bibitem{Bealing2010}
{\sc C.~Bealing}, {\sc R.~Martoň\'{a}k}, and {\sc C.~Molteni},
\newblock {\em Solid State Sciences} {\bf 12}, 157 (2010).

\bibitem{Bochicchio}
{\sc Bochicchio},
\newblock {Study of B1-B2 transition in colloidal cluster}.

\bibitem{Liu2011}
{\sc J.~Liu} and {\sc Q.~Ge},
\newblock {\em Journal of physics. Condensed matter : an Institute of Physics
  journal} {\bf 23}, 345401 (2011).

\bibitem{sprint}
{\sc F.~Pietrucci} and {\sc W.~Andreoni},
\newblock {\em Phys. Rev. Lett.} {\bf 107}, 85504 (2011).

\bibitem{Tribello2010}
{\sc G.~a. Tribello}, {\sc M.~Ceriotti}, and {\sc M.~Parrinello},
\newblock {\em Proc. Natl. Acad. Sci. U. S. A.} {\bf 107}, 17509 (2010).

\bibitem{Santarossa2010}
{\sc G.~Santarossa}, {\sc A.~Vargas}, {\sc M.~Iannuzzi}, and {\sc A.~Baiker},
\newblock {\em Phys. Rev. B} {\bf 81}, 174205 (2010).

\bibitem{Pavan2013}
{\sc L.~Pavan}, {\sc C.~{Di Paola}}, and {\sc F.~Baletto},
\newblock {\em Eur. Phys. J. D} {\bf 67}, 24 (2013).

\bibitem{Lipscomb1966a}
{\sc W.~N. Lipscomb},
\newblock {\em Science (New York, N.Y.)} {\bf 153}, 373 (1966).

\bibitem{Mackay1962}
{\sc A.~L. Mackay},
\newblock {\em Acta Crystallographica} {\bf 15}, 916 (1962).

\bibitem{Paz-Borbon2008a}
{\sc L.~O. Paz-Borb\'{o}n}, {\sc R.~L. Johnston}, {\sc G.~Barcaro}, and {\sc
  A.~Fortunelli},
\newblock {\em The Journal of chemical physics} {\bf 128}, 134517 (2008).

\bibitem{Rosato89}
{\sc V.~Rosato}, {\sc M.~Guillop\'{e}}, and {\sc B.~Legrand},
\newblock {\em Philos. Mag. A} {\bf 59}, 321 (1989).

\bibitem{Cleri1993}
{\sc F.~Cleri} and {\sc V.~Rosato},
\newblock {\em Physical Review B} {\bf 48}, 22 (1993).

\bibitem{Wales2015}
{\sc D.~J. Wales},
\newblock {\em The Journal of Chemical Physics} {\bf 142}, 130901 (2015).

\bibitem{Pietrucci2015}
{\sc F.~Pietrucci} and {\sc R.~Martoň\'{a}k},
\newblock {\em The Journal of chemical physics} {\bf 142}, 104704 (2015).

\bibitem{Uppenbrink1991}
{\sc J.~Uppenbrink} and {\sc D.~J. Wales},
\newblock {\em Journal of the Chemical Society, Faraday Transactions} {\bf 87},
  215 (1991).

\bibitem{Honeycutt1987}
{\sc J.~D. Honeycutt} and {\sc H.~C. Andersen},
\newblock {\em The Journal of Physical Chemistry} {\bf 91}, 4950 (1987).

\bibitem{Chen2008}
{\sc F.~Chen} and {\sc R.~L. Johnston},
\newblock {\em Applied Physics Letters} {\bf 92}, 023112 (2008).

\end{thebibliography}

\end{document}